\documentclass[twocolumn,amsmath,amssymb,aip,10pt]{revtex4-1}
\usepackage{graphicx}% Include figure files
\usepackage{ulem}
\usepackage{hyperref}

\newcommand\paperB[6]{{#1,~}{%
{``#2''},~}{#3~}{{\bf #4},~}{#5~}{(#6)}}

\newcommand\bookB[4]{%
{#1,~}{{\em #2},~}{(#3,~}{#4)}}

\begin{document}
\title{Improving student understanding of electrodynamics: the case for differential forms}
\author{S. Fumeron} 
\email{sebastien.fumeron@univ-lorraine.fr}
\author{B. Berche}
%\author{ T. Bertels} 
\author{ F. Moraes}
\altaffiliation{Permanent address: Departamento de F\'{\i}sica, Universidade Federal Rural de Pernambuco, 
52171-900, Recife, PE, Brazil}
\affiliation{Laboratoire de Physique et Chimie Th\'eoriques,
CNRS UMR 7019, Universit\'e de Lorraine, 54506, Vand\oe{uvre}-les-Nancy, France.}

%************************************************************************************************
\begin{abstract}
The illuminating role of differential forms in electromagnetism is seldom discussed in the classroom. It is the aim of this article to bring forth some of the relevant insights that can be learnt from a differential forms approach to E\&M. The article is self-contained in that no previous knowledge of forms is needed to follow it through. The effective polarization of  the classical vacuum due to a uniform gravitational field and of the quantum vacuum in the Casimir effect are used to illustrate the power and easiness of interpretation of differential forms in dealing with electromagnetism in nontrivial situations. We hope that this article motivates the physics teacher to bring the subject of differential forms to the classroom.
\end{abstract}
%************************************************************************************************
\maketitle
%************************************************************************************************
\section{Introduction}

Why is undergraduate electromagnetism so commonly considered difficult? In an essay entitled \textit{Why is Maxwell's theory so hard to understand?},\cite{Dyson1999} F. Dyson suggests the answer lies in the concept of field, introduced firstly by M. Faraday. Actually, understanding what is the electromagnetic field requires to give up the familiar concepts of Newtonian mechanics (acceleration, mass, force...) in favor of intangible objects remote from directly accessible experience. 

Yet mechanical models and classical electromagnetism (EM) still keep something in common: an extensive use of the concept of three-dimensional vector. Although J.C. Maxwell originally used quaternions algebra in his \textit{Treatise on electricity and magnetism} (1873), J.W. Gibbs, O. Heaviside and H. Hertz developed vector calculus to rewrite Maxwell's equations into the more compact form familiar to every student. In free space, these equations are given in SI units by: \cite{BornWolf, Jackson}  
\begin{eqnarray}
\textbf{curl} \: \mathbf{E}+\frac{\partial \mathbf{B}}{\partial t}&=&0 \label{MaxF} \\
\text{div} \: \mathbf{D}&=&\rho \label{MaxG} \\
\textbf{curl} \: \mathbf{H}-\frac{\partial \mathbf{D}}{\partial t}&=& \mathbf{j} \label{MaxA} \\
\text{div} \: \mathbf{B}&=&0 \label{MaxT} \\
\mathbf{D}&=&\varepsilon_0 \mathbf{E} \label{const1} \\
\mathbf{B}&=&\mu_0 \mathbf{H} \label{const2}
\end{eqnarray}
Here, the source-terms in the nonhomogeneous equations (\ref{MaxG}) and (\ref{MaxA}) are $\rho$, the electric free charge density, and $\mathbf{j}$, the electric free current density. $\mathbf{E}$ is the electric field, $\mathbf{D}$ is the electric flux density, $\mathbf{H}$ is the magnetic field and $\mathbf{B}$ is the magnetic flux density. 

The first four equations are desirable because of Helmholtz decomposition theorem, which states that any vector field (vanishing at infinity) is unique and well-defined provided its curl and its divergence are specified (note that these latter are supplied not only from the free charges and currents, but also from the partial time derivatives of the fields). The last two equations, known as the constitutive relations, are required in order to close the system of equations. 

A foretaste of the misconceptions encountered in EM is obtained by paying attention to the terminology. On the one hand, depending on the context, \textbf{H} is called `magnetic field' (our choice here), but also `magnetic field intensity', `magnetic field strength' or even `magnetising force' (electrical engineering): the latter three suggest that \textbf{H} is not the fundamental magnetic quantity (see for example \cite{Gratus2019}). On the other hand, \textbf{B} is known as `magnetic induction field' (ought to Faraday's law), `magnetic flux density' (our choice here) but EM textbooks sometimes refer to it as the `magnetic field'. \cite{Griffith, Zangwill} The electric part of the field has a less disparate terminology: \textbf{E} is always called `electric field' and only \textbf{D} is known as 'electric flux density', 'electric displacement field' or 'electric induction'. There again, the latter two underline the idea of \textbf{D} being an auxiliary quantity and therefore, we opted for the less-used term of 'electric flux density' (an important asset of such choice is that it emphasizes the parallel between \textbf{E} and \textbf{H} on the one hand, between \textbf{D} and \textbf{B} on the other hand). 

All this may come from a widespread way for building Maxwell's equations in matter (see for example the review \cite{Sipe}): as suggested by Lorentz, electrons and nuclei in a medium produce rapidly varying microscopic fields ${\mathbf e}({\mathbf x},t)$ and ${\mathbf b}({\mathbf x},t)$, that obey Maxwell's equations in free space. Performing a Lorentz-Rosenfeld averaging procedure allows to retrieve macroscopic quantities and in particular, the smoothing of source-terms gives rise to new macroscopic effective fields, namely \textbf{D} and \textbf{H}, which thus appear as auxiliary quantities (or excitations) relevant primarily in matter. Therefore, in vacuum constitutive relations, \textbf{B} and \textbf{H} (resp. \textbf{D} and \textbf{E}) seem to be redundant fields as they are simply equal up to a multiplicative physical constant. 

A particularly insightful exercise is then to examine how the fields transform under a general coordinate transformation $\left\{x,y,z\right\}\rightarrow \left\{u(x,y,z), v(x,y,z), w(x,y,z)\right\}$. Denoting by \textit{J} the Jacobian matrix, Pendry and Ward \cite{Pendry1996} showed that Maxwell's equations remain invariant provided the old and the new fields (primed quantities) are related by:
\begin{eqnarray}
\mathbf{E}'&=&J^{-T} \mathbf{E}\;\;\;\;\mathbf{H}'=J^{-T} \mathbf{H} \label{t1} \\
\mathbf{D}'&=&\frac{J}{\text{det} J} \mathbf{D}\;\;\;\;\mathbf{B}'=\frac{J}{\text{det} J} \mathbf{B} \label{t2}
\end{eqnarray}
\textbf{D} and \textbf{E} (resp. \textbf{B} and \textbf{H}) do not transform in the same way and hence, they are not redundant although mathematically, they are described by the same kind of object: vectors. 

Another intriguing point is that \textbf{E} and \textbf{H} obey a similar transformation law (as \textbf{D} and \textbf{B} do), suggesting the pair share a common mathematical nature. Sometimes, an additional clarification is introduced based on mirror symmetries, dividing the fields into polar (or true) vectors such as \textbf{E} and \textbf{D}, and axial vectors (or pseudovectors) such as \textbf{B} and \textbf{H}.\cite{Jackson} 

However, it is obviously not enough to enlighten the form of all (\ref{t1})-(\ref{t2}) and as remarked by Kitano, \cite{Kitano12} ``\textit{in spite of the simple appearance, the constitutive relations, even for the case of vacuum, are the non-trivial part of the EM theory.}".    

The fact that \textbf{D} and \textbf{E} (resp. \textbf{B} and \textbf{H}) are not redundant in free space is largely unknown and it raises at least two questions: 1) What is the true nature of their connection? 2) Are there pedagogical examples that could help illustrate their different physical contents? In this paper, we use exterior calculus to answer these two questions and in doing so, all the points raised in the previous paragraphs. 
 
Exterior calculus originates from the pioneering works of Grassmann and Cartan and it is concerned with the properties of mathematical objects called differential forms. These latter have raised considerable attention because of their many applications in physics, as testified by the number of articles \cite{Kitano12,Deschamps81,Schleifer83,Hoshino78,Amar80,Baldomir86,Bossavit91,Texeira99-2,Warnick2} and books \cite{MSW,Thirring,Burke,Baez,Flanders,Lindell,frankel2011geometry,Felsager,Bachman,Dray} dedicated to differential forms, but they are usually not introduced at an undergraduate level and as far as we know, only more advanced textbooks make abundant use of this formalism (see e.g. \cite{Thirring,Felsager,Zee,GockelerSchucker}).

In the first section of this work, the basic ideas of exterior calculus are introduced in an intuitive way. Only a very basic knowledge of linear algebra will be used to define differential forms along with the different operations which they allow to perform. 
A particular emphasis will be put on the assets of this formalism from the standpoint of teaching EM in introductory university courses. Then, we illustrate and discuss some key differences between \textbf{D} and \textbf{E} by working out two examples. 

\section{A primer on exterior calculus}
\subsection{What are differential forms?}
Introducing differential forms in a handy but yet accurate way is probably the most challenging part of an exterior-calculus-based electrodynamics course. \cite{Schleifer83} Undergraduate students are generally familiar with exact (or total) differentials, as those abound in thermodynamics courses. Given a Cartesian coordinate system on the Euclidean space $\mathbb{R}^3$, the total differential of any scalar function $F$ writes as (Einstein's summation convention on repeated indices is used)
\begin{equation}
dF=\frac{\partial{F}}{\partial x}dx+\frac{\partial{F}}{\partial y}dy+\frac{\partial{F}}{\partial z}dz=\frac{\partial{F}}{\partial x^a}dx^a . \label{tot-diff}
\end{equation}
When changing the function, only the partial derivatives will change: this means that anytime an exact differential is computed, one is working in a vector space whose basis elements are the $\left\{dx^a, a=1..3\right\}=\left\{dx, dy, dz\right\}$. 

More generally, any object (not necessarily a total differential) that is written as a linear combination of $dx^j$ is called a 1-form and it belongs to the 1-form vector space (or cotangent space) denoted by $\Lambda^1\left(\mathbb{R}^3\right)$. Once given the 1-form vector space, more general objects can be built in a natural way: this is the idea underlying the more general concept of tensors (technically, a (\textit{m},\textit{p})-tensor is simply a multilinear map acting on a collection of \textit{m} 1-forms and \textit{p} vectors to produce a real number - for more details, see for example \cite{Schutz}). Taking the antisymmetrized tensor product (denoted for short by $\wedge$) for each pair of 1-form basis elements gives
\begin{equation}
dx^a \otimes dx^b-dx^b \otimes dx^a=dx^a \wedge dx^b .\label{wedge-1f}
\end{equation}
Here, the regular tensor product $\otimes$ is defined as the ordered product of pairs of 1-forms (and of vectors) and it is associative.

Now, it appears that one can generate only a finite number of non-zero terms, which are a linearly independent and spanning subset of a new vector space: $\Lambda^2\left(\mathbb{R}^3\right)$, the space of 2-forms. For example, in Cartesian coordinates, the 2-form basis written in the right cyclic order is the set: $dy \wedge dz, dz \wedge dx, dx\wedge dy$ (no element $dx^a$ is repeated as (\ref{wedge-1f}) would return 0). That process can be iterated for $p$-uples basis elements $\left\{dx^{a_1}\wedge..\wedge dx^{a_p}, a_1\neq..\neq a_p\right\}$ and generates forms of degree $p$ (or \textit{p}-forms) that belong to a vector space $\Lambda^p\left(\mathbb{R}^3\right)$ of dimension 
$C^p_3=3\:!/(p\:!(3-p)\:!)$. By construction, a \textit{p}-form is a completely antisymmetric (0,$p$) tensor. Concretely, for Cartesian coordinates in $\mathbb{R}^3$, the general expression for a form of degree
\begin{eqnarray}
0\;\;\:\text{is}&&\;\:\;f(x,y,z) \nonumber \\
1\;\:\;\text{is}&&\;\:\;f_1(x,y,z)dx+f_2(x,y,z)dy+f_3(x,y,z) dz \nonumber \\
2\;\;\:\text{is}&&\:\;\;g_1(x,y,z)dy\wedge dz+g_2(x,y,z)dz\wedge dx \nonumber \\
&&\;\;\;\;+g_3(x,y,z) dx\wedge dy \nonumber \\
3\;\;\:\text{is}&&\:\;\;g(x,y,z) dx\wedge dy\wedge dz \nonumber
\end{eqnarray}

As seen above, the cotangent space is a vector space associated to 1-forms. In a similar fashion one associates ordinary vectors to a `tangent space' as follows.  A vector can be seen as an operator which, when acting on a function, gives its directional derivative along the vector direction. That is, taking  the directional derivative of a function $F$ along the vector  $\mathbf{v}$ at the point $x$ one gets
\begin{equation}
   D_{\mathbf{v}} F(x)= \mathbf{v}\cdot \mathbf{\nabla}F(x)=v^a {\partial_a}F(x).
\end{equation}
We see that there is a one-to-one correspondence between vectors and derivations such that the vector $\mathbf{v}$ may be represented by its components in the basis $\{ \partial_a\}$. It follows that the linear space spanned by this basis is a vector space called the tangent space at the point $x \in \mathbb{R}^3$. Like in the case of the \textit{p}-forms, the wedge product between vectors can be used to generate higher order vectors, or \textit{p}-vectors.

The spaces of \textit{p}-forms and of \textit{p}-vectors can be defined on base spaces that are more general than $\mathbb{R}^3$. For any manifold $\mathcal{M}$ of dimension $n$ (that is a smooth hypersurface that locally looks like $\mathbb{R}^n$), once given a coordinate system $\left\{x^a, a=1..n\right\}$, one gets the vector coordinate basis $\left\{\textbf{e}_a=\partial/\partial x^a=\partial_a\right\}$ and a 1-form basis $\left\{dx^a\right\}$. These two sets are said dual, in the sense that when they act upon each other, it returns a real number as prescribed by
\begin{equation}
dx^a\left(\partial_b\right)=\partial_b\left(dx^a\right)=\delta^a_b. \label{ortho}
\end{equation}
In other words, a 1-form is a linear map from a vector space onto $\mathbb{R}$ and more generally, a \textit{p}-form is a multilinear map acting on a collection of  \textit{p}-vectors onto $\mathbb{R}$. Note that in mathematics, there are other notions of duality, such as Hodge duality which will be introduced in section \ref{sec-hodge}. 

Next to that algebraic standpoint on forms, another way to grasp forms is from integral calculus. Indeed, differential forms ``\textit{are the things which occur under integral signs}":\cite{Flanders} a form of degree $p$ is simply an object that one integrates $p$-times to get a scalar. Stated otherwise, ``\textit{a differential form is simply this: an integrand}" .\cite{Bachman} A particular care must be taken when introducing forms that way. 

First, exterior calculus relies strictly speaking on Lebesgue integrals, not on Riemann integrals.

Second, it is customary in calculus to drop the wedge sign when computing double (multiple) integrals such as $\iint_R f(x,y) dx dy$, where the domain of integration is defined as $R=[x_1,x_2]\times[y_1,y_2]$. But then, the Fubini's exchange-of-integration-order formula may look contradictory with the antisymmetry property of the 2-form $dx\wedge dy$. Although subtle, the disagreement is nothing but apparent: indeed, switching $dx$ with $dy$ also involves a reparametrization of the integration domain $p:R\rightarrow R'=[y_1,y_2]\times[x_1,x_2]$. As for any $p$-form \textit{u} $\iint_R u=-\iint_{p(R)} u$ if the map $p$ reverses the orientation of the integration domain (conversely, there is a plus sign if the map preserves orientation), then the antisymmetry feature of forms is retained in the customary calculus formulas. 

Third, introductory calculus courses define integrals as the limits of Riemann sums when the partition becomes infinitely fine, $\int^{b}_{a} f(x) dx=\underset{\Delta x \to 0}{\lim} \underset{k}{\sum} f(x_k) \Delta x$, where the $x_k$ are points evenly spaced in the interval $\left[a,b\right]$. This suggests that ``$dx$" is in practice a shorthand notation for an infinitely thin width, which it is not. As explained by Spivak,\cite{Spivak} ``\textit{Classical differential geometers (and classical analysts) did not hesitate to talk about `infinitely small' changes $dx^i$ of the coordinates $x^i$, just as Leibnitz had. No one wanted to admit that this was nonsense, because true results were obtained when these infinitely small quantities were divided into each other}". A rigorous connection between infinitesimal elements and 1-forms belongs to the realm of non-standard analysis, a branch of mathematics developed in the 60s by Abraham Robinson.\cite{Robinson} 

To sum up, the differential form $dx$ is rightly understood as a basis element of the cotangent space, but not as a tiny change: this is why one favors the introduction of differential forms from the algebraic standpoint instead of the more widespread analysis standpoint.   
 
A popular way to gain insight on 1, 2 and 3-forms in $\mathbb{R}^3$ is to use graphical representations known as Schouten pictograms.\cite{Obukhov} The main idea is to plot a \textit{p}-form field $\omega$ as a foliation built from its kernel $\text{Ker}\:\omega$. Recalling that differential forms act on vectors returning  real numbers (see Eq. (\ref{ortho})), the kernel of a form $\alpha$ is the set of all vectors $\mathbf{v}$ such that $\alpha(\mathbf{v})=0$. A foliation is a family of hypersurfaces of dimension $3-p$, filling $\mathbb{R}^3$ with no overlap. One can define it from the kernel if and only if $\text{Ker}\:\omega$ is an integrable plane field (this is known as Pfaff's problem and a general criterion involving the exterior derivative $d$ introduced in the next paragraph is $\omega \wedge d\omega=0$). For 1-forms, integrability means that there is always a 2-surface that is locally tangent to the plane field defined by $\text{Ker}\:\omega$. To be more concrete, let us focus on the 1-form $\omega=dr$ in cylindrical coordinates $\left\{r,\theta,z\right\}$: its kernel is simply obtained at each point from (\ref{ortho}) $dr(\partial_{\theta})=0, dr(\partial_z)=0$. Hence, the plane field generated by vectors $\left\{\partial_{\theta}, \partial_z\right\}$ is the family of planes, tangent to 2-surfaces $r=C^{st}$. 

Although it might look appealing, this visual interpretation
may help for a first initiation to the language of forms, but this rarely goes beyond an introductory level: as discussed in Ref. \onlinecite{Bachman}, there are many simple forms whose kernels are not integrable and present ``contact structures" (for example $xdy+dz$). Hence, we will not rely on that point of view in the reminder of this work.

\subsection{Wedge product, exterior derivative and vector calculus identities}
Relation (\ref{wedge-1f}) introduces an algebraic operation between 1-forms: the wedge (or exterior) product, denoted by the symbol $\wedge$. More generally, given two forms $u, v$ (of degree \textit{m}), $w$ (of degree \textit{p}) and $z$ (of degree \textit{q}), it can be shown easily that the wedge product obeys the three following properties:
\begin{eqnarray}
&& u\wedge w=(-1)^{mp} w\wedge u \;\;\;\;\;\;\text{(supercommutativity)} \label{wp-p1} \\
&& u\wedge\left(w\wedge z\right)=\left(u\wedge w\right)\wedge z \;\;\;\;\;\;\;\;\;\;\;\;\text{(associativity)} \label{wp-p2} \\
&& (\alpha u+\beta v)\wedge w=\alpha u\wedge w+\beta v\wedge w \:\;\;\;\;\text{(linearity)} \label{wp-p3}
\end{eqnarray}
$u\wedge w$ is a form of degree $m+p$. As remarked previously, supercommutativity implies that $dx^a\wedge dx^a=0$: hence, for a base space of dimension $n$, no form of degree $p>n$ can exist. The set of all $\left\{\Lambda^p\left(\mathcal{M}\right), p=0..n\right\}$ equipped with the exterior product defines an algebra known as the Grassmann algebra.

Although the meaning of the wedge product may be obscure, it is still a familiar object to students in the usual Euclidean space. 

When $u$ is a 0-form (a scalar) and $w$ is a \textit{p}-form, the wedge is simply the scalar multiplication and it will be omitted. 

When computing the wedge product of two 1-forms $u=u_a dx^a$ and $v=v_a dx^a$, it comes
\begin{eqnarray}
u\wedge v=&&\left(u_y v_z-u_z v_y\right)dy \wedge dz+\left(u_z v_x-u_x v_z\right)dz \wedge dx \nonumber \\
&&+\left(u_x v_y-u_y v_x\right)dx \wedge dy. \label{ex-cross}
\end{eqnarray}
The components of $u\wedge v$ look like the components obtained from the usual cross product between vectors $\left(u_x \mathbf{e}_x+u_y \mathbf{e}_y+u_z \mathbf{e}_z\right)\times\left(v_x \mathbf{e}_x+v_y \mathbf{e}_y+v_z \mathbf{e}_z\right)$ (the exact forms-vectors connection and its inferences will be refined later, see also Ref. \onlinecite{Hoshino78}). Roughly speaking, when applied to 1-forms in $\mathbb{R}^3$, $\wedge$ has the meaning of a cross-product. 

What about the wedge product between a 1-form and a 2-form? A straightforward calculation shows that for a 1-form $u$ and a 2-form $w$ $=w_x dy\wedge dz + w_y dz\wedge dx + w_z dx \wedge dy$, the wedge product yields
\begin{equation}
u\wedge w=\left(u_x w_x+u_y w_y+u_z w_z\right) dx\wedge dy \wedge dz. \label{ex-ps}
\end{equation}
This time, $\wedge$ behaves like the ordinary dot product between the components of $u$ and $w$.

The wedge product concatenates two forms to produce a new form of higher degree. Another possibility to increase the degree of a form is to take the exterior derivative of a form. The exterior derivative of a \textit{p}-form is a linear mapping from $\Lambda^p\left(\mathcal{M}\right)$ to $\Lambda^{p+1}\left(\mathcal{M}\right)$ defined formally in \textit{n} dimensions as
\begin{equation}
d\omega\equiv\left(\frac{\partial}{\partial x^a}dx^a\right)\wedge \omega\;\;\;\;\;\;\;\;a=1..n \label{deriv-def}
\end{equation}
It generalizes the notion of total differential of a scalar function (\ref{tot-diff}). Unlike the ordinary derivative, it is a dimensionless operator. Besides, it obeys the two following properties
\begin{eqnarray}
&&d\left(u\wedge v\right)=\left(du\right)\wedge v+\left(-1\right)^p u\wedge dv \;\text{(Leibniz)} \label{deriv-p1} \\
&&d\left(du\right)=0 \;\;\;\;\;\;\;\;\;\;\;\;\;\;\;\;\;\;\;\;\;\;\;\;\;\;\;\;\;\;\;\;\;\;\;\;\;\;\;\;\text{(Nilpotence)} \label{deriv-p2}
\end{eqnarray}
with \textit{p} the degree of \textit{u}. 

From the nilpotence property, if a $p$-form $\omega$ is exact (i.e. there is a $(p-1)$-form $u$ such that $\omega=du$), then it is necessarily closed (its exterior derivative vanishes). Conversely, Poincar\'e's lemma states that a closed \textit{p}-form is always locally exact but generally not globally, depending on the topology of the base space $\mathcal{M}$ (this is the object of the de Rham cohomology). 

The exterior derivative of terms such as $g\left(x^1,x^2...\right) dx^{a_1}\wedge..\wedge dx^{a_p}$ is obtained using both the Leibniz formula and the nilpotence property in (\ref{deriv-def})
\begin{equation}
d\left(g\left(x^1,x^2...\right) dx^{a_1}\wedge..\wedge dx^{a_p}\right)=\left(dg\right)\wedge  dx^{a_1}\wedge..\wedge dx^{a_p} \label{prac-deriv-def}
\end{equation}
In practice, the exterior derivative of a \textit{p}-form simply consists in computing the sum of the partial derivatives of its components and then in discarding terms within which a same $dx^a$ is repeated. 

It is interesting to point out the coherence of the notations employed heretofore: for the 0-form $u=x^1$, (\ref{prac-deriv-def}) leads to $du=d(x^1)=dx^1$. 

The exterior derivative must not be confused with either the Lie derivative (which computes variations with respect to a vector field and is connected to $d$ via the Cartan formula) or with the covariant derivative (which is dedicated to parallel transport and requires a connection on the manifold).

Similarly to the wedge product, some insight on $d$ can be gained by expressing (\ref{deriv-p1})-(\ref{deriv-p2}) with forms in $\mathbb{R}^3$. For example in Cartesian coordinates, the exterior derivative of a 0-form $F\left(x,y,z\right)$ gives (\ref{tot-diff}), but now with a clarified meaning for the $dx^a$. Hence, the exterior derivative of a 0-form returns a 1-form with components similar to the ordinary gradient operator. Likewise, the exterior derivative of a 1-form $u$ gives after straightforward algebra
\begin{eqnarray}
du=&&\left(\partial_x u_y-\partial_y u_x\right)dx\wedge dy+\left(\partial_y u_z-\partial_z u_y\right)dy\wedge dz \nonumber \\
&&+\left(\partial_z u_x-\partial_x u_z\right)dz\wedge dx . \label{ex-curl} 
\end{eqnarray}
One recognizes a 2-form whose components correspond to the ordinary curl operator of a vector field with appropriate components (discussed later). Finally, applying the exterior derivative to a 2-form $w$ leads to a 3-form with components analogous to the ordinary divergence operator
\begin{eqnarray}
dw=\left(\partial_x w_x+\partial_y w_y+\partial_z w_z\right)dx\wedge dy\wedge dz . \label{ex-div}
\end{eqnarray}

To sum up, grad, curl and div are just a unique operator in disguise, $d$, applied to forms of different degrees. But there is more: the exterior derivative not only unifies grad, curl and div, but it also generalizes them to arbitrary dimensions (this is responsible for Burke's thought-provoking statement ``Div, Grad, Curl are Dead".\cite{Burke-draft})

As a sidenote, it is worth noting that the inverse operation to derivation, integration, also exists in exterior calculus, but there are some subtleties compared to ordinary integration familiar to students. There is indeed a natural way to integrate a $p$-form over a $p$-dimensional submanifold (note that the degree of the form has to be the same as the dimension of the domain of integration), which does not rely on lengths, angles or scalar products and is hence purely topological. Such integral is not defined in the sense of Riemann but it is properly performed with respect to Lebesgue measure.\cite{Choquet-Bruhat} In practice, this means that integration of a 1-form $u$ on a path $\mathcal{C}$
\begin{equation}
\int_{\mathcal{C}} u=\int_{\mathcal{C}} u_x dx+u_y dy+ u_z dz \nonumber
\end{equation}
can not be interpreted as the usual circulation of a vector field $\mathbf{u}$ along the curve $\mathcal{C}$
\begin{equation}
\int_{\mathcal{C}} \mathbf{u}.\mathbf{ds} \nonumber
\end{equation}
as the scalar product requires the knowledge of the geometry of $\mathcal{M}$ (technically, the metric). But there is worse: there is no consistent way to define the flux of $u$, as that would involve integration of a 1-form on a domain of dimension 2. As can be guessed, the degree of $u$ must be increased beforehand, as expressed by the generalized Stokes theorem:\cite{Amar80}

\begin{equation}
\int_{\partial\mathcal{D}} u = \int_{\mathcal{D}} du. \label{stokes-thm}
\end{equation}

\subsection{Hodge star operator}\label{sec-hodge}
Another ingredient is necessary to reformulate EM with differential forms. To apprehend this, consider the example of a vector field whose counterpart is a 1-form field. As we just saw, the exterior derivation only allows us to build a curl-like equation, leaving the divergence part undefined. This violates Helmholtz theorem (whatever its time-dependent form may be \cite{Heras16}), meaning that the field is not defined unambiguously. Another variation on the same theme: in virtue of the nilpotence property (\ref{deriv-p2}), having only $\wedge$ and $d$ at our disposal leaves no room for equations involving a Laplacian. As we shall see, these two problems are solved by the Hodge duality, which consists in a mapping between \textit{p}-forms and ($n-p$)-forms (remember that $n$ is the dimension of the manifold). Beforehand, one must introduce two objects: the metric tensor and the inner product.

So far, all mathematical objects that have been introduced are purely topological, i.e. they are relying neither on angles nor on distances. These latter can be computed only when the base space is equipped with a metric structure ($\mathcal{M}$ is then called a pseudo-Riemannian manifold). The metric consists in a symmetric $\left(0,2\right)$ tensor denoted by $g$, which defines the scalar product between two vectors:

\begin{equation} 
g\left(\mathbf{u},\mathbf{v}\right)=g_{ab}\:dx^a\left(\mathbf{u}\right)\otimes dx^b\left(\mathbf{v}\right).
\end{equation}

From (\ref{ortho}), its appears that the components of $g$ are simply the scalar products between the coordinate basis elements $g_{ab}=g\left(\partial_a,\partial_b\right)$. What is the physical content of these components? In the flat Eulidean plane, the displacement vector $\mathbf{\Delta s}=\left(\Delta x, \Delta y\right)$ has a length squared given by Pythagoras theorem $\Delta s^2=\Delta x^2+\Delta y^2$, so that $g_{11}=1, g_{22}=1, g_{12}=g_{21}=0$. However, on  the surface of a sphere (of radius $R$), an arc of length squared is given by $\Delta s^2=R^2\Delta \theta^2+R^2 \sin^2\theta \Delta \phi^2$ so that $g_{11}=R^2, g_{22}=R^2 \sin^2\theta, g_{12}=g_{21}=0$. From these examples, one understands that the components of $g$ are related in particular to the curvature of the manifold, a property which remains true in higher dimensions. The metric is real and symmetric, hence the inverse metric exists and obeys $g_{ac}g^{cb}=\delta_a^{\;\;b}$. This provides a one-to-one correspondence between the components of a 1-form $v$ and those of a vector $\mathbf{v}$:
\begin{equation}
v_a=g_{ab} v^b . \label{connection-1fv}
\end{equation}

Locally, there is always an orthonormal coordinate basis such that $g_{ab}$ reduces to a diagonal matrix having only $+1$'s or $-1$'s on its diagonal: the signature of the metric is the couple $(s,n-s)$ where $s$ is the number of -1's. From the standpoint of general relativity, the metric tensor is Lorentzian (e.g. with signature=(1,3)) and it is obtained as a solution of Einstein's field equations.\cite{MSW, Schutz, Carroll, Price} 

The inverse metric $g^{ab}$ defines the inner product $\left\langle ,\right\rangle$ between two \textit{p}-forms according to:
\begin{eqnarray}
&&p=1:\;\;\left\langle dx^a, dx^b \right\rangle=g^{ab} \\
&&p>1:\;\;\left\langle dx^{a_1}\wedge..\wedge dx^{a_p}, dx^{b_1}\wedge..\wedge dx^{b_p} \right\rangle \nonumber \\
&&\;\;\;\;\;\;\;\;\;\;\;\;\;\;\;=\left|\begin{pmatrix}
g^{a_1 b_1} & \cdots & g^{a_1 b_p}\\
\vdots&\ddots&\vdots\\
g^{a_p b_1}&\cdots&g^{a_p b_p}
\end{pmatrix}\right| .
\end{eqnarray}
The inner product between 1-forms is the equivalent of the dot product for vectors. 

Finally, having at our disposal the metric and the inner product, the Hodge dual operator (or star operator $\star$)
is an invertible linear map between $v\in\Lambda^p\left(\mathcal{M}\right)$ and $\star\: v \in \Lambda^{n-p}\left(\mathcal{M}\right)$ such that \cite{Baez}
\begin{equation}
u\wedge\left(\star\: v\right)=\left\langle u,v \right\rangle \sqrt{\left|\text{det}g_{ab}\right|}\: dx^1\wedge..\wedge dx^n \label{Hodge-def}
\end{equation}
(here $u$ is of the same degree as $v$). 

An equivalent and more explicit definition is given by \cite{gasperini2013theory}
\begin{equation}
    \star v = \frac{1}{(n-p)!}v^{\mu_1...\mu_p}\sqrt{|\text{det}g_{ab}|}\epsilon_{\mu_1...\mu_n}dx^{\mu_{p+1}}\wedge ... \wedge dx^{\mu_{n}}, \label{Hodge-better}
\end{equation}
where 
\begin{equation}
    v=v_{\mu_1...\mu_p}dx^{\mu_{1}}\wedge ... \wedge dx^{\mu_{p}}.
\end{equation}
and the totally anti-symmetric Levi-Civita symbol is defined as
\begin{eqnarray}
   \epsilon_{\mu_1...\mu_n}&=&+1\;\;\text{if $\mu_1,..\mu_n$ is an even permutation of 1,..n} \nonumber \\
   &=&-1\;\;\text{if $\mu_1,..\mu_n$ is an odd permutation of 1,..n} \nonumber \\
   &=&0\;\;\text{otherwise}.
\end{eqnarray}
Contrary to $\wedge$ and $d$, the Hodge dual is a metric-dependent operator and it obeys the properties:
\begin{eqnarray}
&&\star \left(\star\: v\right)=\left(-1\right)^{s+p(n-p)}v \label{hodge-id} \\
&& u\wedge \left(\star\:v\right)=v\wedge \left(\star\:u\right)  \label{hodge-com} \\
&& u\wedge \left(\star\:u\right)=0\Rightarrow u=0 .
\end{eqnarray}

To understand what (\ref{Hodge-def}) really does, let us illustrate its action on the $p$-form basis for the flat Euclidean space in spherical coordinates $\left\{r,\theta,\phi\right\}$:
\begin{eqnarray}
&&\star\left(d\theta\wedge d\phi\right)=\frac{1}{r^2 \sin\theta}dr, \\ &&\star\left(dr\wedge d\theta\right)=\sin\theta d\phi  \\
&&\star\left(d\phi\wedge dr\right)=\frac{1}{\sin\theta}d\theta, \\ &&\star\left(dr\wedge d\theta\wedge d\phi\right)=\frac{1}{r^2 \sin\theta}. 
\end{eqnarray}

These relations are of course completed by means of (\ref{hodge-id}). When the metric is diagonal, they are obtained for any curvilinear coordinates by following the simple recipe: respecting the right cyclic order, the action of the Hodge dual on a $p$-uple returns the missing ($n-p$)-uple, each $dx^a$ being multiplied by its corresponding Lam\'e coefficient (the square root of the metric $\sqrt{g_{aa}}$ (no index summation)). Although $\star\:v$ appears as some kind of orthogonal complement of the \textit{p}-form $v$, we will see later that in EM it can  be pictured more accurately as a rotation between the electric and magnetic parts of the field.  

Having the Hodge operator at our disposal, one is now in position to address the questions asked at the beginning of this section. Symmetrically to the exterior derivative which increases the degree of forms, one can now define a derivation operation that lowers the degree of the forms: this is the coderivative $\delta$ defined as
\begin{equation}
\delta v= \left(-1\right)^{s+p(n-p)}\star d\star v \label{def-coder}.
\end{equation}
The coderivative maps $\Lambda^p\left(\mathcal{M}\right)\rightarrow \Lambda^{p-1}\left(\mathcal{M}\right)$, in a similar fashion as a divergence operator which returns a scalar from a higher-degree object, vector or pseudovector. It obeys the two following properties:\cite{Fecko}
\begin{eqnarray}
&& \delta (\delta u)=0 \;\;\;\;\;\;\;\;\;\;\;\;\;\;\;\;\;\;\text{(Nilpotence)} \label{cod-p1} \\
&& du\wedge (\star\: v)=u\wedge \star (\delta v)+d\left(u\wedge \star v\:\right) \label{cod-p2}
\end{eqnarray}
That is the missing piece required to construct second derivatives that do not identically vanish. Indeed, the anticommutator between $d$ and $\delta$, known as the Laplace-de Rham operator, generalizes the usual Laplace operator for vectors in $\mathbb{R}^3$ to \textit{p}-forms on an arbitrary base space $\mathcal{M}$:
\begin{equation}
\Delta v=\left(\delta d+d\delta \right) v \label{def-Laplace}.
\end{equation}
It also provides a way to close the set of equations required to define the field: instead of the curl and divergence required for vector fields, form fields are unambiguously defined by their derivative and their coderivative. 

This is the main result of the Hodge decomposition theorem, that generalizes the Helmholtz theorem.\cite{Kobe86} Moreover, the Hodge operator connects in a precise manner $\wedge$, $d$ and their counterparts in vector analysis. The case of 1-forms has already been settled: they are translated unambiguously into vectors as prescribed by (\ref{connection-1fv}). Now, looking back at (\ref{ex-ps}), we remarked that when dealing with a 1-form $u$ and a 2-form $v$, $u\wedge v$ behaves like the ordinary dot product: but as the right hand side is a 3-form and not a 0-form (a scalar), it is in fact $\star (u \wedge v)$ (and not simply $u \wedge v$) that actually returns a 0-form and corresponds to the ordinary dot product. A similar statement holds for divergence which corresponds to $\star\: d$ of a 2-form (and not simply $d$) as appears from (\ref{ex-div}). More generally, taking the Hodge dual translates unambiguously 3-forms into scalars. The case  of 2-forms may look trickier at first glance, as they can be translated either as bivectors (using the metric two times to raise each index) or as vectors (taking the Hodge star and then using the metric on the 1-form obtained). However, it can be shown after some algebra that, for a diagonal metric in 3 dimensions, these two procedures lead to the same components (this is no longer true in 4 dimensions). Hence, as summarized in Ref. \onlinecite{Schleifer83}, for $u,v\in \Lambda^1{\mathcal{M}}$, $\textbf{u}\times \textbf{v}$ identifies with $\star (u \wedge v)$ and {\bf curl} \textbf{u} with $\star\: d u$.

\section{Outcomes}
\subsection{Unification of vector analysis}
As seen previously, depending on the degree of the forms between which it is applied, and up to a Hodge dual, the wedge product unifies scalar multiplication, cross product and scalar product whereas exterior derivative unifies grad, curl and div operators. Bearing that in mind, let us investigate on a few examples how (\ref{deriv-p1})-(\ref{deriv-p2}) are translated for forms of different degrees in $\mathbb{R}^3$. Beginning with the nilpotence property applied to a 0-form $F$, $dF$ stands for the gradient and as it is a 1-form, its exterior derivative  gives the curl, so that in the end one retrieves the well-known formula
\begin{equation}
\star\:d\left(dF\right)=0\;\rightarrow\;\mathbf{curl}\left(\mathbf{grad}\: F\right)=0.
\end{equation}
For a 1-form $u$, the same approach leads to
\begin{equation}
\star\:d\left(du\right)=0\;\rightarrow\; \text{div}\left(\mathbf{curl}\: \mathbf{u}\right)=0.
\end{equation}
Considering the Leibniz formula, in the case of two 0-forms $F$ and $G$, it comes that
\begin{eqnarray}
&&d\left(F\wedge G\right)=\left(dF\right)\wedge G+\left(-1\right)^0 F\wedge dG \nonumber \\
&&\;\rightarrow\;\mathbf{grad}\left(F\:G\right)=G\:\mathbf{grad}\: F+F\: \mathbf{grad}\; G.
\end{eqnarray}
For $F$ a 0-form and $u$ a 1-form, one obtains
\begin{eqnarray}
&&\star\:d\left(F\wedge u\right)=\star\left(dF\wedge u\right)+\left(-1\right)^0 F \left(\star\:du\right) \nonumber \\
&&\;\rightarrow\;\mathbf{curl}\left(F\:\mathbf{u}\right)=\mathbf{grad}\: F\times \mathbf{u}+F \mathbf{curl}\; \mathbf{u}.
\end{eqnarray}
For two 1-forms, this now gives
\begin{eqnarray}
&&\star\:d\left(u\wedge v\right)=\star\left(du\wedge v\right)+\left(-1\right)^1 \star\left(u\wedge dv\right) \nonumber \\
&&\;\rightarrow\;\text{div}\left(\mathbf{u}\times \mathbf{v}\right)=\left(\mathbf{rot}\: \mathbf{u}\right).\mathbf{v}-\mathbf{u}.\left(\mathbf{rot}\: \mathbf{v}\right).
\end{eqnarray}

As can be understood from these examples, the various identities of vector analysis simply come down to two short formulas, (\ref{deriv-p1}) and (\ref{deriv-p2}). Other identities involving the mixed product and triple vector product can also be derived from the antisymmetry properties of the wedge product.\cite{Schleifer83}
Finally, all formulas involving integration of scalar and vector fields can also be retrieved from (\ref{stokes-thm}). Indeed, when applied to a 0-form $F(x)$, it simply gives the fundamental theorem of calculus
\begin{eqnarray}
\int_{\left[a,b\right]} dF=\int_{\partial\left[a,b\right]} F \;\rightarrow\; \int_{\left[a,b\right]} \frac{\partial F}{dx^i}dx^i=F\left(b\right)-F\left(a\right). \nonumber
\end{eqnarray}
Once a metric is given, one can draw connections between the usual integral identities and the generalized Stokes theorem: the Stokes formula is obtained from (\ref{stokes-thm}) applied to a 1-form, whereas the Gauss formula comes from (\ref{stokes-thm}) applied to a 2-form. The Green identities are obtained by integrating (\ref{cod-p2}) and then using the Stokes theorem \cite{Fecko}
\begin{eqnarray}
&&\int_{\mathcal{D}} dF\wedge (\star dG)+\int_{\mathcal{D}} F\wedge (\star \Delta G)=\int_{\partial\mathcal{D}} F \wedge \star dG \\
&&\int_{\mathcal{D}} F (\star \Delta G)-G (\star \Delta F)=\int_{\partial\mathcal{D}} F (\star d G)-G (\star d F)\;\;\;\;
\end{eqnarray}
for $F$ and $G$ two 0-forms. Integration of the Leibniz formula also provides an interesting result:
\begin{eqnarray}
\int_{\mathcal{D}}\left(du\right)\wedge v=\int_{\partial\mathcal{D}} u\wedge v-\left(-1\right)^p \int_{\mathcal{D}} u\wedge dv
\end{eqnarray}
where (\ref{stokes-thm}) was used to simplify the second integral: this expression generalizes integration by parts at arbitrary dimensions. 
\subsection{``Forms illuminate EM..."}
In this section, we reformulate electromagnetic theory in the language of differential forms and in so doing we discuss its axiomatics. Indeed, there are some redundancies within the set of equations, as can be seen from the Maxwell-Thomson equation (\ref{MaxT}). In virtue of Schwarz's theorem, $\text{div}\:\mathbf{B}=0\:\Rightarrow\:\text{div}\: \partial_t\mathbf{B}=0$, so that $\partial_t \mathbf{B}$ is a purely solenoidal field. Hence, it expresses as the curl of a ``potential", which is nothing else than Maxwell-Faraday equation (\ref{MaxF}). In order to clarify the foundations and the structure of EM, we therefore proceed by trying to use the minimal set of mathematical and physical objects at each step: in particular, following the prescription of Obukhov and Hehl, \cite{Obukhov} we will try to postpone the use of the metric (which originates from another interaction, gravitation) to as late as possible.
Let us begin with a very general statement: electric charge is a quantifiable physical property. In practice, given a 3-dimensional compact domain $\mathcal{V}$, it means that one can always count (at least classically) the number of elementary charges localized within $\mathcal{V}$ and then obtain the total charge $Q$. This defines a 3-form electric charge density $\rho$ such that
\begin{equation}
Q=\int_{\mathcal{V}}\rho \label{def-charge}.
\end{equation}
This relation is purely topological, as it does not rely on lengths or angles. As $\rho$ is a 3-form in $\mathbb{R}^3$, then necessarily
\begin{equation}
d\rho=0 .
\end{equation}
This equation has no analog in the usual Maxwell-Heaviside formulation of EM and comes from the fact that electric charge is quantifiable. A first outcome arises in virtue of the Poincar\'e lemma: as $\rho$ is a closed form, it is locally exact so that there exist a potential 2-form $D$ such that
\begin{equation}
\rho=dD \label{MG-forms}.
\end{equation}
One recognizes here the exterior calculus version of the Maxwell-Gauss equation (\ref{MaxG}). Hence, electric flux density translates into a 2-form $D$. 

A second outcome appears when taking the time derivative of $\rho$ and using Schwarz's theorem: as $d(\partial_t \rho)=0$, in virtue of the Poincar\'e lemma again, $\partial_t \rho$ is exact so that there exists in any contractible domain a 2-form potential $j$ such that
\begin{equation}
\partial_t \rho=d(-j) \label{CE-forms}.
\end{equation}
One recognizes here the continuity equation for the electric charge, which is usually built as a spin-off of (\ref{MaxG}) and (\ref{MaxA}). Note that as suggested by the usual continuity equation, the term in $\text{div}\:\mathbf{j}$ confirms that the electric current density is indeed a 2-form. But there is more: combining (\ref{MG-forms}) and (\ref{CE-forms}), it comes that $d\left(j+ \partial_t D\right)=0$, so that once again from the Poincar\'e lemma, there exists a 1-form potential $H$ such that
\begin{equation}
j+ \partial_t D=dH \label{MA-forms}.
\end{equation}
This is the Maxwell-Amp\`ere equation (\ref{MaxA}). 

To summarize, \textit{D} and \textit{H} are potentials associated to sources $\rho$ and $\mathbf{j}$ and they are introduced from topological equations, which result from the fact that electric charges are quantifiable. Hence, they are relevant not only in matter, but also in free space and they ``are microphysical quantities of the same type likewise - in contrast to what is stated is most textbooks".\cite{Obukhov2} 

The last set of equations is based on the absence of magnetic monopoles, which can also be expressed from a counting procedure on the 3D compact domain $\mathcal{V}$ (also coming from the idea that charges are quantifiable):
\begin{equation}
Q_M=0=\int_{\mathcal{V}}\rho_M ,\label{mag-charge}
\end{equation}
so that the 3-form magnetic charge density identically vanishes, $\rho_M=0$. This is the physical content of the Maxwell-Thomson equation (\ref{MaxT})
\begin{equation}
dB=0, \label{MT-forms}
\end{equation}
where $B$ is the 2-form magnetic flux density. Hence, from the Poincar\'e lemma, once again deriving (\ref{MT-forms}) with respect to time and using the Schwarz's theorem, there exists a 1-form potential $E$, the electric field, such that
\begin{equation}
\partial_t B=d(-E). \label{MF-forms}
\end{equation}
This is the exterior algebra version of the Maxwell-Faraday equation (\ref{MaxF}) and its similarity with (\ref{CE-forms}) highlights its status of continuity equation for the magnetic flux. As suggested by the usual form of the Maxwell-Faraday equation, the term in $\mathbf{curl}\:\mathbf{E}$ confirms that the electric field is indeed a 1-form. 

Moreover, forms and their associated vector fields do not share the same dimension. Combining (\ref{def-charge}) with (\ref{MG-forms}) and using the Stokes theorem gives the Gauss theorem
\begin{equation}
Q=\int_{\partial\mathcal{V}} D .\label{Gauss-thm}
\end{equation}
A quick dimensional analysis reveals that the dimension of $D$ is that of a charge $[D]={\tt C}$ (but its components recover the regular dimension ${\tt C}/{\tt m}^2$), $[j]={\tt C}/{\tt s}$ $[H]={\tt C}/{\tt s}$, $[B]=flux={\tt J.s}/{\tt C}$ and $[E]={\tt J}/{\tt C}$.
We are now in position to address the problem of the transformation laws (\ref{t1})-(\ref{t2}) mentioned in the introduction. Indeed, consider a general coordinate change $\left\{x,y,z\right\} \rightarrow \left\{u\left(x,y,z\right), v\left(x,y,z\right), w\left(x,y,z\right)\right\}$ , one is looking for the components of $E$ in the new coordinate system. Straightforward algebra gives
\begin{eqnarray}  
E&=&E_a dx^a=E_x\left(\frac{\partial x}{\partial u}du+\frac{\partial x}{\partial v}dv+\frac{\partial x}{\partial w}dw\right)+... \nonumber \\
&=&\left(\frac{\partial x}{\partial u}E_x+\frac{\partial y}{\partial u}E_y+\frac{\partial z}{\partial u}E_z\right)du+... \nonumber \\
&=&\left[\left(J^{-T}\right)_{11}E_x+\left(J^{-T}\right)_{12}E_y+\left(J^{-T}\right)_{13}E_z\right]du+...\nonumber,
\end{eqnarray}
which agrees with (\ref{t1}). On the other hand, for the 2-form $D$, it comes that
\begin{eqnarray}  
D&&=D_x dy\wedge dz+D_y dz\wedge dx+D_z dx\wedge dy \nonumber \\
&&=D_x \left(\frac{\partial y}{\partial u^a}du^a\right)\wedge\left(\frac{\partial z}{\partial u^b}du^b\right) +... \nonumber \\
&&=D_x\left[\left(\frac{\partial y}{\partial v}\frac{\partial z}{\partial w}-\frac{\partial z}{\partial v}\frac{\partial y}{\partial w}\right)dv\wedge dw+...\right]+... \nonumber \\
&&=D_x\left[\left(\text{Com}\:J^{-1}\right)_{11}dv\wedge dw+\left(\text{Com}\:J^{-1}\right)_{12}dw\wedge du\right. \nonumber \\
&&\;\left.+\left(\text{Com}\:J^{-1}\right)_{13}du\wedge dv...\right]+... \nonumber \\
&&=\left[D_x\left(\frac{J}{\text{det J}}\right)_{11}+D_y\left(\frac{J}{\text{det J}}\right)_{21}+D_z\left(\frac{J}{\text{det J}}\right)_{31}\right] \nonumber \\
&&dv\wedge dw+... \nonumber,
\end{eqnarray}
which agrees with (\ref{t2}). The notation $\text{Com}\: M$ refers to the matrix of cofactors of $M$. The transformation laws for $\mathbf{E}$ and $\mathbf{D}$ are different because, fundamentally, these correspond to two different mathematical objects: forms of different degrees.

Now, considering (\ref{MG-forms}), (\ref{MA-forms}), (\ref{MT-forms}) and (\ref{MF-forms}), it appears that the number of unknowns exceeds the number of equations. The role devoted to the constitutive relationships is precisely to fill this gap by providing additional connections between the different parts of the electromagnetic field (as the purpose of the present work is mainly pedagogical, we will not discuss the more technical cases of non-linear media and magneto-electric effects). As in 3D 1-forms and 2-forms are naturally mapped into each other by the Hodge dual operator, the only admissible translation of (\ref{const1})-(\ref{const2}) is:
\begin{eqnarray}
D&=&\varepsilon_0 \star E \label{forms-c1}, \\
B&=&\mu_0 \star H \label{forms-c2} .
\end{eqnarray}
The constitutive relations are metric-coupled dual relations needed to close Maxwell's equations. This highlights one extremely important point: as the four Maxwell's equations can be built without the metric, it is the main purpose of constitutive relations to reveal how the EM field couples to geometry. Moreover, in free space, it is customary from the standpoint of vector analysis to proceed to Gauss field identification $\mathbf{E}=\mathbf{D},\; \mathbf{H}=\mathbf{B}$ (up to dimensional multiplicative constants). As mentioned in section II.C, when the metric is diagonal, this identity applies but anytime the metric has off-diagonal terms, this identification fails (see for instance the mixing of EM components in the Sagnac effect.\cite{Post67}) Writing constitutive relations from exterior algebra clarifies why Gauss identification may not apply: the action of Hodge operator - and consequently the constitutive properties of vacuum - results from the geometric properties of space, so that anytime these latter are non-trivial, the links between the different fields are non-trivial as well (in 4D, vacuum constitutive properties are only related to the conformal part of the metric \cite{Guillemin}). 

In other words, vacuum has to be considered as a full-fledged medium with its own optical properties, as testified by the terminology of $\varepsilon_0$, i.e. the permittivity of free space. Like in any material medium, the electric part of the field is determined by both $\mathbf{E}$ and $\mathbf{D}$ (resp. by $\mathbf{H}$ and $\mathbf{B}$ for the magnetic part) that encompass information of different kinds and are not redundant quantities even in free space. This is in fact very close to Maxwell's original view on the field, as expressed explicitly in his \textit{Treatise}: ``\textit{We are thus led to consider two different quantities, the magnetic force and the magnetic induction, both of which are supposed to be observed in a space from which the magnetic matter is removed.}".\cite{Maxwell}

\subsection{``... and EM illuminates forms."\cite{MSW}}
The formula defining the Hodge dual makes its meaning rather abstruse. Although $\star\:u$ might appear as some kind of orthogonal complement of the \textit{p}-form $u$, a deeper way to understand Hodge star is related to an internal symmetry of Maxwell's equations, known as S-duality.

In the absence of sources, these latter are left invariant when performing ``cross-rotations" between the electric and magnetic parts of the field: \cite{Jackson, Zangwill}
\begin{eqnarray}
\mathbf{E}'&=&\mathbf{E} \cos\alpha -\mathbf{B} \sin\alpha \\
\mathbf{B}'&=&\mathbf{E} \sin\alpha+\mathbf{B} \cos\alpha
\end{eqnarray}
where for simplicity we use units for which $c=1$. Following a remark by Weber,\cite{Weber01} this duality rotation can be expressed in a more compact form from the Weber vector $\mathbf{F}=\mathbf{E}+i\mathbf{B}$ (cf. Ref. \onlinecite{Birula13}):
\begin{eqnarray}
\mathbf{F}'&=&e^{i\alpha}\:\mathbf{F}=\left(\cos\alpha+i\sin\alpha\right)\left(\mathbf{E}+i\mathbf{B}\right) \\
&=&\left(\mathbf{E}\cos\alpha - \mathbf{B}\sin\alpha\right)+i\left(\mathbf{E}\sin\alpha+\mathbf{B}\cos\alpha\right). \label{rot-RS}
\end{eqnarray}
$\alpha$ is a mixing angle related to the relative proportions of electric and magnetic fields.

A direct connection with the Hodge operator can be made but at the cost of working in four dimensions. In this case, the electric 1-form $E$ and the magnetic 2-form $B$ are gathered within the Faraday 2-form 
\begin{equation}
F=E\wedge dt+B    . \label{Farad}
\end{equation}
Taking the 4D-exterior derivative and using (\ref{MF-forms}) gives:
\begin{eqnarray}
d_4 F&=&dE\wedge dt+dB+\partial_t B\wedge dt \nonumber \\
&=&(-\partial_t B)\wedge dt+\partial_t B\wedge dt=0.
\end{eqnarray}
Similarly, one defines the Maxwell 2-form as
\begin{equation}
G=\star_4 F=D-H\wedge dt \label{Max-2f}
\end{equation}
and this latter obeys the source equation:
\begin{eqnarray}
d_4 G&=&dD+\partial_t D \wedge dt -dH\wedge dt \nonumber \\
&=&\rho+\partial_t D \wedge dt -\left(j+dD\right)\wedge dt \nonumber \\
&=&\rho-j\wedge dt=J.
\end{eqnarray}

When applied on a 2-form in a 4D Lorentzian metric, (\ref{hodge-id}) becomes $\star_4^2=-1$, in a similar fashion to the imaginary unit $i^2=-1$. Hence, one constructs the following operator on a 2-form as
\begin{equation}
\exp\left(\star_4\: \alpha\right)=\cos\alpha +\star_4\:\sin\alpha \label{dual-rot}.
\end{equation}
Like the complex number $\exp(i\theta)$, it obeys
\begin{eqnarray}
&&e^{\star_4\: \alpha}e^{\star_4\: \beta}=e^{\star_4\: \left(\alpha+\beta\right)}=e^{\star_4\: \beta}e^{\star_4\: \alpha} \\
&&e^{\star_4 \frac{\pi}{2}}=\star_4.
\end{eqnarray}

In Cartesian coordinates, applying the dual rotation operator (\ref{dual-rot}) to the Faraday 2-form leads to
\begin{eqnarray}
e^{\star_4\: \alpha}F&=&\cos\alpha\: F +\sin\alpha \star_4 F \nonumber \\
&=&\cos\alpha\left(E_x dx\wedge dt+..+B_x dy\wedge dz+..\right) \nonumber \\
&+&\sin\alpha\star_4\left(E_x dx\wedge dt+..+B_x dy\wedge dz+..\right) \nonumber \\
&=&\left(E_x \cos\alpha-B_x \sin\alpha\right) dx\wedge dt+.. \nonumber \\
&+&\left(E_x \sin\alpha+B_x \cos\alpha\right)dy\wedge dz+..
\end{eqnarray}
A comparison with (\ref{rot-RS}) shows that the Hodge operator acts as a dual rotation between electric 2-forms and magnetic 2-forms. In the next section, we explore how the Hodge dual applies on the EM field in the presence of non trivial vacuua.

\section{Vacuum as a polarizable medium}

\subsection{Classical vacuum: electrostatics in a uniform gravitational field}

As it is well known, Newton's law of gravitation as given by the potential $\Phi (\boldsymbol{r})$, may be obtained from any metric theory of gravitation which, in the weak field limit, turns into 
\begin{equation}
    ds^2=-\left( 1+{2\Phi} \right)dt^2 + d\ell^2 , \label{Newtonian}
\end{equation}
where $d\ell$ represents the space part of the metric.

Whittaker \cite{whittaker1927electric} took the  limit of Schwarzschild's metric, 
\begin{equation}
    ds^2=-\left( 1-\frac{2M}{r} \right)dt^2 +\frac{1}{1-\frac{2M}{r}}dr^2 +r^2\left( d\theta^2 +\sin^2\theta d\varphi^2  \right) ,
\end{equation}
at a large distance from the gravitating centre  such that  the Newtonian potential $-gz$ appears in the time part of the metric.  The result is
\begin{equation}
    ds^2=-\left( 1+{2gz} \right)dt^2 +\left( dx^2 +dy^2 +\frac{dz^2}{1+{2gz}} \right),  \label{uniform}
\end{equation}
which thus corresponds to an attractive uniform Newtonian gravitational field parallel to the $z$-axis.

Considering an electric field weak enough  such that its contribution to the gravitational field of this spacetime is negligible,  Whittaker derived Laplace's equation for the scalar electrostatic potential in the absence of sources:
\begin{equation}
\frac{\partial^2 V}{\partial x^2} + \frac{\partial^2 V}{\partial y^2} + \left( 1+{2gz} \right) \frac{\partial^2 V}{\partial z^2}    =0 . \label{Laplace}
\end{equation}

Now, one considers a uniformly charged infinite $x$-$y$ plane. Denoting by $\sigma$ the uniform areal charge density, the corresponding volume charge density 3-form is $\rho=\sigma\delta (z)\, dx\wedge dy \wedge dz$. It can be easily seen that the displacement field 2-form
\begin{equation}
  D=\frac{1}{2}[\Theta(z) -\Theta(-z)]\sigma \, dx\wedge dy  \label{D-form}
\end{equation}
 obeys the Maxwell-Gauss law $dD=\rho$ and the expected planar symmetry $D(-z)=-D(z)$. Here  $\Theta(z)$ is the Heaviside step function.

Now, the 2-form $D$ is related to $E$ via $D=\varepsilon_0 \star_4 (E\wedge dt)$ and symmetry tells us that the electric field 1-form writes $E= \Tilde{E}\, dz$. We use (\ref{Hodge-better}) to compute $\star_4 (dz \wedge dt)=dx\wedge dy$. Note that  $|\text{det}g_{ab}|=1$ for the metric (\ref{uniform}). We then have
\begin{equation}
D= \epsilon_0  \Tilde{E}dx \wedge dy 
\end{equation}
and finally that
\begin{equation}
  E= \frac{1}{2\varepsilon_0}[\Theta(z) -\Theta(-z)]\sigma   \, dz . \label{E-form}
\end{equation}

On the other hand, Eq. (\ref{Laplace}) becomes $\left( 1+{2gz} \right)\frac{\partial^2 V}{\partial z^2}=0$, or $\frac{\partial^2 V}{\partial z^2}=0$, which gives the usual $g$-independent 0-form potential
\begin{equation}
    V(x,y,z)=-\frac{\sigma}{2\epsilon_0} |z| . \label{pot}
\end{equation}
With $E=-dV$ one confirms the electric field 1-form given by Eq. (\ref{E-form}). 

Comparing the expressions (\ref{D-form}) and (\ref{E-form}) reveals that $D_{xy}=\varepsilon_0 E_z$, similarly to what happens in flat spacetime: hence, there is no vacuum polarization due to the uniform gravitational field.

\subsection{Quantum vacuum: wave propagation in a Casimir vacuum} 

The Casimir effect was first discovered by Hendrik Casimir in 1948 and it corresponds to a macroscopic force due the quantum vacuum fluctuations. That force appears when two parallel plates are separated from one another by a vacuum layer of thickness $a$ and it is generally attractive. Propagation of an electromagnetic field in the so-called Casimir vacuum can be adequately described as experiencing the following geometry: \cite{Visser}
\begin{equation}
g_{\mu\nu}=\eta_{\mu\nu}-\frac{d_2(z)}{d_1(z)+d_2(z)}n_{\mu}n_{\nu}
\end{equation}
where $\eta_{\mu\nu}=(-1,+1,+1,+1)$ is the flat Minkowski metric and $n_{\mu}=(0,0,0,1)$ is the unit vector orthogonal to the plates. The variable $z$ is the coordinate $x^{3}$, which is  measured in the direction perpendicular to the plates. Functions $d_1(z)$ and $d_2(z)$ are determined from quantum electrodynamics and a perturbative expansion up to order $\alpha^2$ gives:\cite{Visser}
\begin{eqnarray}
d_1(z)&=&-\frac{1}{16\pi}+O(\alpha^2) \\
d_2(z)&=&-\frac{11\pi\alpha^2}{64800\: a^4 m_e^4}
\end{eqnarray}
with $m_e$ the electron mass. In Cartesian coordinates, the Casimir vacuum metric can thus be approximated by
\begin{equation}
    g_{\mu\nu}=
    \begin{bmatrix}
-1 & 0 & 0 & 0 \\
0 & 1 & 0 & 0 \\
0 & 0 & 1 & 0 \\
0 & 0 & 0 & 1-C\alpha^2
\end{bmatrix} \label{CV}
\end{equation}
where $C=11\pi^2/4050\:a^4 m_e^4$. In the absence of magnetic sources, Eq. (\ref{Max-2f}) becomes
$D=\star_4 F$. Now, using (\ref{Hodge-better}) we have for the components of $D$
\begin{equation}
 D_{\rho\sigma}=\frac{1}{2}F^{\mu\nu}\sqrt{|g|}\epsilon_{\mu\nu\rho\sigma}= \frac{1}{2}g^{\mu\alpha}g^{\nu\beta}F_{\alpha\beta}\sqrt{|g|}\epsilon_{\mu\nu\rho\sigma}  \label{Dcomps}  .
\end{equation}
This leads to
\begin{equation}
    D^j=\sqrt{-g}|g^{00}| g^{jj} E_j
\end{equation}
with no summation on the $j$ index. Ought to the symmetries of the problem, only the
z-component of the field is relevant and therefore using (\ref{CV}) leads to
\begin{eqnarray}
\varepsilon=1-\frac{11\pi^2}{8100\:a^4 m_e^4}\alpha^2+O(\alpha^2) \label{eps-c}
\end{eqnarray}
Although the second term is probably far too small to be detectable, quantum vacuum behaves in principle as a polarized medium, as is well-known in quantum electrodynamics. As there are no gradient of the dielectric constant, the refractive index of the Casimir vacuum is a constant given by the square root of (\ref{eps-c}). Note that as emphasized in Ref.  \onlinecite{Visser}, the dielectric constant being the same along the three axes, no birefringence is expected.

\section{Concluding remarks}

In the wake of Ref. \onlinecite{Warnick1}, we hold exterior calculus to be a precious pedagogical tool for teaching electrodynamics. Although students are more comfortable with vector analysis, picturing the EM field as three-dimensional vectors turns out to be misleading in some cases. Conversely, picturing the EM field as differential forms is probably less intuitive at first sight, but it leads to a deeper understanding of EM.

The main concepts of exterior calculus were introduced from undergraduate linear algebra and calculus. Contrary to a widespread misconception, differential forms are not tiny quantities, they are dual vectors. The many formulas used in vector analysis can be concisely written in terms of the wedge product and the exterior derivative. Exterior calculus also helps clarifying the axiomatics of EM. From two postulates on electric charge (Eq. (\ref{def-charge})) and magnetic charge (Eq. (\ref{mag-charge})) or equivalently (\ref{MT-forms}), the whole set of Maxwell's equations can be recovered without relying on the background geometry: in other words, their foundation is purely topological and does not require any metric structure. Metric properties of the manifold appear on the other hand in the constitutive relations. \textbf{D} and \textbf{E} (resp. \textbf{H} and \textbf{B}) must indeed be understood as forms of different degrees that carry different information. They are connected by a duality relation that requires the knowledge of the background geometry, as the Hodge star operator is explicitly depending on the metric.

According to Sir W.L. Bragg, ``the important thing about science is not so much to obtain new facts as to discover new ways of thinking about them" (cited in Ref. \onlinecite{Bragg}). For such a purpose, the relevance of exterior calculus goes far beyond the realm of electrodynamics. As a matter of fact, this formalism is likely to enlighten conceptual issues arising in thermodynamics, fluid mechanics...\cite{Flanders} Forthcoming developments may also involve the Yang-Mills theories: these are generalizations of Maxwell's equations but built from larger symmetry groups ($SU(N)$ instead of $U(1)$) and they require additional ingredients taken from exterior calculus. 

\begin{acknowledgments}
The authors thank T. Bertels for insightful discussions. FM wishes to acknowledge the support of LPCT/U.Lorraine and CNPq.
\end{acknowledgments}

\end{document}